\documentclass[english,3p,authoryear]{elsarticle}
\usepackage{mathpazo}
\usepackage[T1]{fontenc}
\usepackage[latin9]{inputenc}
\usepackage{amsmath}
\usepackage{graphicx}
\usepackage{setspace}
\makeatletter
\providecommand{\tabularnewline}{\\}
\journal{Sustainability}
\makeatother
\usepackage{babel}
\begin{document}
	\begin{frontmatter}{}
		
		\title{Newton--Raphson Emulation Network for Highly Efficient Computation of Numerous Implied Volatilities}
		
		\author[labela]{Geon Lee}
		\author[labelb]{Tae-Kyoung Kim}
		\author[labelc]{Hyun-Gyoon Kim}
		\author[labeld]{Jeonggyu Huh\corref{cor1}}
		\address[labela]{Department of Mathematics \& Statistics, Chonnam National University, Gwangju 61186, Korea}
		\address[labelb]{50 Yeouinaru-ro, Yeongdeungpo-gu, Seoul 07328, Republic of Korea}
		\address[labelc]{School of Mathematics \& Computing, Yonsei University, Seoul 03722, Korea}
		\address[labeld]{Department of Statistics, Chonnam National University, Gwangju 61186, Korea}
		\cortext[cor1]{corresponding author.\\
			Email address: huhjeonggyu@jnu.ac.kr}
		\begin{abstract}
			In finance, implied volatility is an important indicator that reflects the market situation immediately. Many practitioners estimate volatility using iteration methods, such as the Newton--Raphson (NR) method. However, if numerous implied volatilities must be computed frequently, the iteration methods easily reach the processing speed limit. Therefore, we emulate the NR method as a network using PyTorch, a well-known deep learning package, and optimize the network further using TensorRT, a package for optimizing deep learning models. Comparing the optimized emulation method with the NR function in SciPy, a popular implementation of the NR method, we demonstrate that the emulation network is up to 1,000 times faster than the benchmark function.
		\end{abstract}
		\begin{keyword}
			Graphics processing unit (GPU) accelerated computing; implied volatility; Newton--Raphson method; PyTorch; TensorRT 
		\end{keyword}
	\end{frontmatter}{}
	
	\section{Introduction}
	Volatility is the degree of variability in underlying asset dynamics, helping investors predict future market variability, and is usually divided into historical and implied volatility. Because historical volatility is obtained from information for a specific period in the past, the type of volatility lags behind the market situation. Unlike historical volatility, implied volatility contains only current market information, not past market information \citep{gatheral2011volatility}. When a sudden shock, such as a financial crisis, occurs, implied volatility is fairly important in predicting future volatility.
	
	When using implied volatility for various purposes, such as estimating parameters of an option pricing model, it is often necessary to convert a large number of option prices into implied volatilities in real time. However, iterative methods, such as the bisection and Newton--Raphson (NR) methods, typically used to obtain implied volatilities, are unsuitable for calculating numerous implied volatilities due to excessive computation. Therefore, many studies \citep{jackel2006implication,mininni2021challenges,orlando2017review,stefanica2017explicit} have proposed several formulas to approximate implied volatility. The implied volatility estimated using the formulas can be more accurately corrected using iterative methods, such as the NR method (J�ckel, P. 2015).
	
	However, increasing the accuracy using various mathematical methods already faces a limitation. Thus, in line with various studies \citep{berg2019data,chen2018neural,li2020fourier,raissi2018hidden,raissi2019physics,ramuhalli2005finite} that have supplemented numerical schemes, such as the finite element method, with neural networks, \citet{liu2019pricing} and \citet{kim2022large} introduced neural networks to improve the accuracy of estimating implied volatility. Although the time required for estimating implied volatility can be greatly reduced through these previous studies, the computation time must be reduced further, considering the case in which numerous implied volatilities must be estimated repeatedly.
	
	In this study, we develop a graphics processing unit (GPU) acceleration scheme for the NR method, reducing the computational time for estimating implied volatilities. To this end, we apply the so-called neural emulation technique, which implements an algorithm as if a neural network with zero or very few parameters. This technique enables employing well-known deep learning packages, such as TensorFlow and PyTorch, to accelerate a scientific procedure. These popular packages make it straightforward to implement large-scale parallel computation using GPUs. Additionally, this approach allows a neural network optimization engine, TensorRT, to further maximize inference performance. We refer to the network emulating the NR method as the NR emulation network.
	
	The presented NR emulation network was compared with the NR method of SciPy, a widely used implementation of the NR method, in terms of estimation accuracy and speed to verify the effectiveness of this study. The test results reveal that the NR emulation network is up to 1,000 times faster than the NR method of SciPy, but with similar accuracy.
	
	The background, such as the implied volatility and the NR method, is provided in the next section. Section 3 fully describes the NR emulation network. The NR network is compared in terms of accuracy and computation time with the benchmark in Section 4. The last section concludes the work.
	
	\section{Background}
	\subsection{Implied Volatility}
	An option is a contract that trades the right to buy (call option) and sell (put option) an asset at a predetermined strike price on a maturity date. In addition, options can be divided into several types depending on the exercise method. If the option can be exercised only on the expiration date of the contract, it is a European-style option. The Black--Scholes model \citep{black1973pricing} is generally used to evaluate European options.
	
	In the Black--Scholes model, the option pricing formula is given by
	\begin{gather}
		c^{call}(S_{t},t;r,\sigma,K,T)=S_{t}N(d_{1})-Ke^{-r(T-t)}N(d_{2})\nonumber \\
		c^{put}(S_{t},t;r,\sigma,K,T)=Ke^{-r(T-t)}N(-d_{2})-S_{t}N(-d_{1}),\label{eq:BS_formula}
	\end{gather}
	where $S_{t}$ is the stock price at $t$, $r$ denotes the risk-free rate, $\sigma$ represents the volatility of $S_{t}$, $K$ and $T$ are the strike price and expiration time of the option, respectively, $d_{1}=\frac{1}{\sigma\sqrt{T-t}}\{\ln\frac{S_{t}}{K}+(r+\frac{1}{2}\sigma^{2})(T-t)\}$, $d_{2}=d_{1}-\sigma\sqrt{T-t}$, $N(\cdot)$ denotes the cumulative distribution function of the standard normal distribution, and $c^{call}$ and $c^{put}$ indicate the prices for the call and put options, respectively. Among the variables that influence the option price, except for the volatility $\sigma$, the other variables $S_{t}$, $t$, $r$, $K$, and $T$ can be provided from the market information and the option specification, whereas $\sigma$ must be estimated using market data to calculate $c_{mkt}$. However, in many cases, the market price $c_{mkt}$ of the option is a known quote because most options are exchange-traded products, and the corresponding $\sigma$ is reversely calculated from the price $c_{mkt}$. The value of $\sigma$ computed in this way is called implied volatility $\sigma_{impl}$.
	
	In other words, for a given $S_{t}$, $r$, $t$, $K$, $T$, and $c_{mkt}$, the implied volatility $\sigma_{impl}$ is defined for each option as follows:
	\begin{equation}
		c_{mkt}=h_{r,k,\tau}(\sigma_{impl}):=c(S_{t},t;r,\sigma_{impl},K,T),
		\label{eq:nonlinear_equation}
	\end{equation}
	where $k=S_{t}/K$ and $\tau=T-t$. Because $h_{r,k,\tau}(\cdot)$ is monotonically increasing, $\sigma_{impl}$ uniquely exists as $h_{r,k,\tau}^{-1}(c_{mkt})$ if $c_{mkt}\in(0,S_{t})$. In addition, $\sigma_{impl}$ is often considered an alternative indicator of $c_{mkt}$ because $\sigma_{impl}$ changes in a more stable way than $c_{mkt}$.
	
	\subsection{Newton--Raphson Iterative Method}
	The nonlinear equation (\ref{eq:nonlinear_equation}) must be solved with a numerical scheme for determining $\sigma_{impl}$ because $h_{r,k,\tau}^{-1}$ is not found explicitly. An iterative method, such as the bisection or secant method, is commonly used to determine a solution to a nonlinear equation. Particularly, the NR method, which is an algorithm with a fast convergence rate, is most used for estimating $\sigma_{impl}$.
	
	According to the NR method, the implied volatility $\sigma_{impl}$ can be obtained in a series of the following update steps:
	\begin{equation}
		\sigma_{n+1}=\sigma_{n}-\frac{h_{r,k,\tau}(\sigma_{n})-c_{mkt}}{h_{r,k,\tau}\prime(\sigma_{n})}.
		\label{eq:NR_step}
	\end{equation}
	If the initial value $\sigma_{0}$ is given within the convergence interval, the NR method converges rapidly to $\sigma_{impl}$ with a quadratic convergence rate. However, there is a risk of divergence if $\sigma_{0}$ is not given in the convergence interval. Fortunately, the convergence of the NR method is guaranteed if $\sigma_{0}$ is set to $\sigma_{c}$, as follows \citep[refer to][]{higham2004introduction}:
	\begin{equation}
		\sigma_{c}=\sqrt{\left|\frac{2}{\tau}(\ln k+r\tau)\right|}\label{eq:initial_point},
	\end{equation}
	where $\sigma_{c}$ is the unique inflection point of $h_{r,k,\tau}$, where the option vomma is 0. The first and second derivatives $\frac{\partial c}{\partial\sigma}$ and $\frac{\partial^{2}c}{\partial\sigma^{2}}$ of the option price $c$ with respect to $\sigma$ are called vega $\upsilon$ and vomma $\upsilon'$, respectively.
	
	\section{Newton--Raphson Emulation Network}
	This section proposes and describes the NR emulation network emulating the NR method. The emulation networkIt enables to obtain numerous implied volatilities in real time through parallel computing of the GPU and optimizing the computation graphs of the network. 
	
	\begin{figure}[h]
		\begin{singlespace}
			\centering{}\includegraphics[scale=0.5]{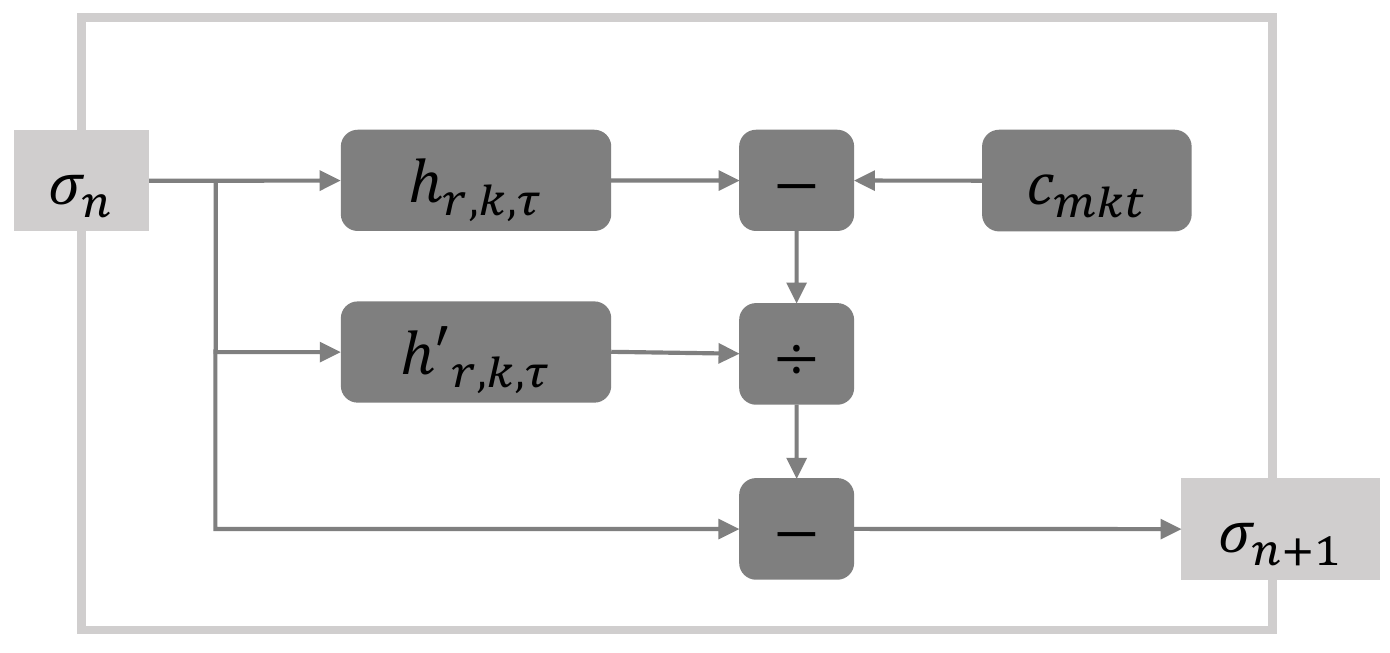}\caption{Newton--Raphson update layer.
				\label{fig:NRU_layer}}
		\end{singlespace}
	\end{figure}
	
	The NR update (NRU) layer depicted in Figure~\ref{fig:NRU_layer} is designed to emulate the update step (\ref{eq:NR_step}) of the NR method. In addition, $h_{r,k,\tau}$ of the NRU layer is defined in (\ref{eq:nonlinear_equation}). Therefore, if the input $\sigma_{n}$ passes through the NRU layer, one step of the NR method is applied to produce $\sigma_{n+1}$, which is expected to be closer to $\sigma_{impl}$ than $\sigma_{n}$. Additionally, $h_{r,k,\tau}$ and $h_{r,k,\tau}'$ depend on the risk-free rate $r$ implicitly, the ratio $k$ of the stock price to the strike price, and the time to maturity $\tau$; thus, the NRU layer also depends on $r$, $k$, and $\tau$. In addition, the NRU layer is also dependent on $c_{mkt}$. This dependence can also be considered for the NRU layer to be conditioned on $r$, $k$, $\tau$, and $c_{mkt}$, similar to the conditional generative adversarial network~ \citep{mirza2014conditional}.
	
	\begin{figure}[h]
		\centering{}\includegraphics[scale=0.5]{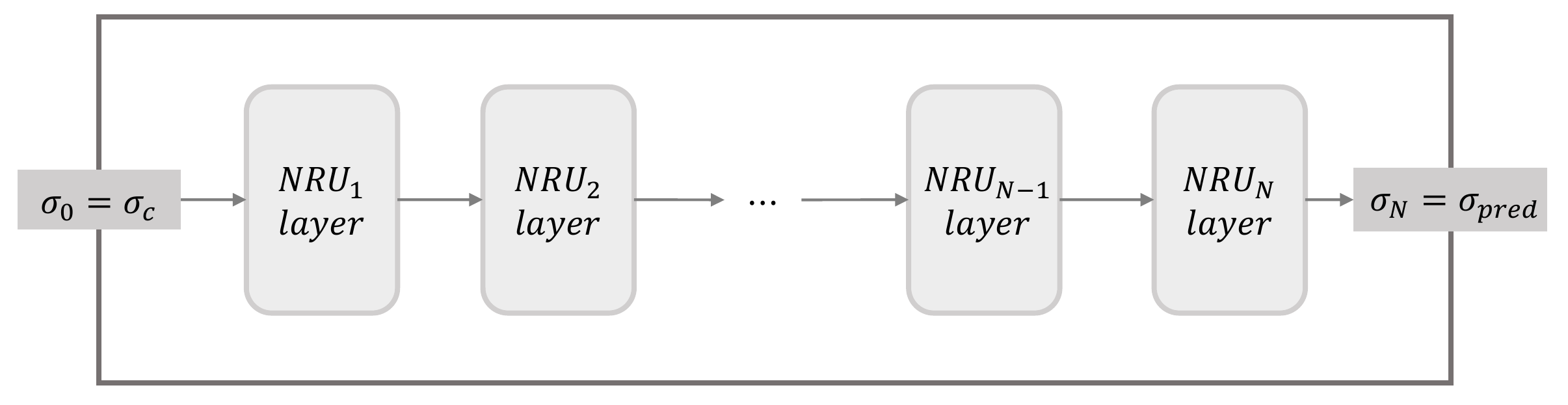}\caption{Newton--Raphson (NR) emulation network. 
			\label{fig:NR_net}}
	\end{figure}
	
	The NR emulation network is created by stacking NRU layers as depicted in Figure~\ref{fig:NR_net}, which corresponds to the process of repeating the update steps of the NR method. As the input $\sigma_{0}$ for the network, $\sigma_{c}$ in Equation~(\ref{eq:initial_point}) is chosen. This choice ensures that the output $\sigma_{pred}$ is sufficiently close to $\sigma_{impl}$ if the emulation network is deep enough.\footnote{Except in the cases where a too-small $\sigma_{impl}$ makes $\sigma_{pred}$ diverge because of the limitations of the floating point number system.} Passing through the deep network means performing the update steps of the NR method many times. In the experiments that follow in the next section, it is empirically demonstrated that the minimum depth of the NR emulation network should be eight to guarantee convergence. In other words, when $\sigma_{0}=\sigma_{c}$, there should be at least eight NRU layers in the network such that $|\sigma_{pred}-\sigma_{impl}|<\epsilon$ for the machine epsilon $\epsilon$($\approx10^{-6}$) of the single-precision floating system.
	
	To exploit powerful parallel computing, we implement the NR emulation network with PyTorch, a well-known deep learning framework, and run it on the GPU. This approach also allows optimizing the network with TensorRT to accelerate the inference performance of the emulation network. In addition, TensorRT is one of the deep learning-related tools provided by NVIDIA, which can be used to optimize the structure of a network while converting a dynamic graph of PyTorch into a static graph.\footnote{https://developer.nvidia.com/tensorrt} Although it is usual to use neural networks to identify patterns inherent in data, such a data-learning stage does not exist in this study.
	
	In the next section, we experimentally reveal how accurately and quickly the NR emulation network derives the implied volatility. We conclude that market prices $c_{mkt}$ of numerous options can be converted into implied volatilities $\sigma_{impl}$ in real time.
	
	\section{Numerical Tests}
	\subsection{Test Data Description}
	A testing dataset with one million data points was prepared by generating virtual option prices $c_{mkt}$ using the Black--Scholes formula (\ref{eq:BS_formula}), converting them to the corresponding implied volatilities $\sigma_{impl}$. The variables $\sigma$, $\tau$, and $k$ involved in generating $c_{mkt}$ are randomly selected within the ranges as in Table~\ref{tab:RV_ranges} (for convenience, the risk-free rate $r$ is fixed to 0 to offset its effect). The variables $\sigma$, $\tau$, and $k$ are the volatility parameter for the Black--Scholes model, time to maturity $T-t$, and ratio $S_{t}/K$ of the stock price to exercise price, respectively. 
	
	\begin{table}[h]
		\centering{}%
		\begin{tabular}{cccc}
			\hline 
			Variable & $\sigma_{impl}$ & $\tau$ & $\ln k$\tabularnewline
			\hline 
			\hline 
			Distribution & $U(0.01,0.5)$ & $U(0.01,2)$ & $U(-\frac{\sigma^{2}}{2}\tau-2\sigma\sqrt{\tau},-\frac{\sigma^{2}}{2}\tau+2\sigma\sqrt{\tau})$\tabularnewline
			\hline 
		\end{tabular}\caption{\label{tab:RV_ranges} Variable ranges involved in generating
			the virtual test data; $U(a, b)$ is the uniform distribution on $(a, b)$.}
	\end{table}
	
	We set the variable ranges to be as acceptable as possible by considering and reflecting the real market. Most options in the real market have a time to maturity $\tau$ of less than two years, and typically, the volatility $\sigma$ does not fall below 1\% and does not exceed 50\%. Moreover, the strike price $k$ is set to be within the 99\% confidence interval of the distribution of the stock price $S_{\tau}$ at time $\tau$, and the distribution is obtained from the assumption of Black and Scholes that $\ln S_{\tau}$ follows $N(-\frac{\sigma^{2}}{2}\tau, \sigma^{2}\tau)$ when $S_{0}=1$ and $r=0$.
	
	\subsection{Test Results}
	In this section, we analyze the results of various tests. In practice, the NR method (SciPy-NR) of the Python package SciPy is often used to obtain implied volatility.\footnote{https://scipy.org/} Therefore, we compared the accuracy and efficiency of the SciPy-NR method and the NR emulation network. The SciPy-NR method estimates the implied volatility through eight iterations, and the emulation network performs the estimation through eight NRU layers. Eight is the minimum number for both methods to reduce errors to near the machine epsilon $\epsilon$ ($\approx10^{-6}$) of the single-precision floating number system. 
	
	\begin{table}[h]
		\begin{centering}
			{\small{}}%
			\begin{tabular}{ccc}
				\hline 
				& SciPy-NR & NR emulation\tabularnewline
				\hline 
				\hline 
				Platform & SciPy (Python) & PyTorch + TensorRT (Python)\tabularnewline
				Hardware & CPU (Intel Xeon Silver 4216) & GPU (NVIDIA GeForce RTX 2080)\tabularnewline
				\hline 
			\end{tabular}{\small\par}
			\par\end{centering}
		\centering{}\caption{\label{tab:plaform}Implementation platform and hardware.}
	\end{table}
	Table \ref{tab:plaform} reveals that the NR emulation network runs on the GPU to take full advantage of parallel computing. However, except for expensive Tesla GPUs, ordinary GPUs specialize in single-precision floating numbers, not double-precision. In this study, we do not have a Tesla GPU; thus, we process the tests based on the single-precision floating number system. Therefore, for a fair comparison, the SciPy-NR method is also conducted with the precision of the single-precision floating numbers.
	
	\begin{table}[h]
		\begin{centering}
			\begin{tabular}{ccc}
				\hline 
				Error type & SciPy-NR & NR emulation\tabularnewline
				\hline 
				\hline 
				MAE & 2.800171e-08 & 2.816055e-07\tabularnewline
				MSE & 1.930116e-15 & 2.949284e-13\tabularnewline
				MRE & 2.155739e-07 & 1.962279e-06\tabularnewline
				\hline 
			\end{tabular}
			\par\end{centering}
		\centering{}\caption{\label{tab:errors}Implied volatility estimation error.}
	\end{table}
	
	Table \ref{tab:errors} compares the accuracy of each method using the mean absolute error (MAE), mean square error (MSE), and mean relative error (MRE) for inferring the implied volatility. The definitions of MAE, MSE, and MRE are provided as follows:
	\begin{gather*}
		{\rm MAE}=\frac{1}{L}\sum_{i=1}^{L}\vert\sigma_{i,pred}-\sigma_{i,impl}\vert,\;{\rm MSE}=\frac{1}{L}\sum_{i=1}^{L}(\sigma_{i,pred}-\sigma_{i,impl})^{2},\;{\rm MRE}=\frac{1}{L}\sum_{i=1}^{L}\frac{\vert\sigma_{i,pred}-\sigma_{i,impl}\vert}{\sigma_{i,impl}},
	\end{gather*}
	where $L=1,000,000$, and $\sigma_{i,pred}$ denotes the value derived by the emulation network to predict $\sigma_{i,impl}$. Both methods achieve the maximum possible accuracy on the single-precision floating number system, as the values of MAE and MRE are below $\epsilon$, and the value of MSE is below $\epsilon^{2}$. The SciPy-NR method tends to infer $\sigma_{impl}$ 10 times more accurately than the emulation network, implying that the CPU may achieve higher precision than the GPU, even if both processing units work on similar single-precision floating number systems.
	
	\begin{table}[h]
		\begin{centering}
			\begin{tabular}{ccc}
				\hline 
				\# of implied volatility estimates & SciPy-NR & NR emulation\tabularnewline
				\hline 
				\hline 
				10,000 & 14.71 (3.0527) & 0.4 (0.0182)\tabularnewline
				100,000 & 99.07 (4.7911) & 0.44 (0.01)\tabularnewline
				1,000,000 & 1212.64 (11.2775) & 1.50 (0.0065)\tabularnewline
				\hline 
			\end{tabular}
			\par\end{centering}
		\centering{}\caption{\label{tab:computation_times}Computation times (in milliseconds) for estimating the implied volatility. Each value is calculated by averaging the values from 100 repetitions, and the corresponding standard deviation is provided in parentheses.}
	\end{table}
	
	Table~\ref{tab:computation_times} presents the computation time consumed for the execution of each method. The NR emulation network has a very short computation time compared to the SciPy-NR method. Additionally, as the number of implied volatility estimates increases, the emulation network becomes overwhelmed by the SciPy-NR method in terms of processing speed. When the number of implied volatility estimates reaches one million, the running time of the network is about 1,000 times shorter than that of the SciPy-NR method. The computation times are repeatably measured 100 times, and the average and standard deviation of the resultant values are written together.
	
	\begin{figure}[h]
		\centering{}\includegraphics[scale=0.35]{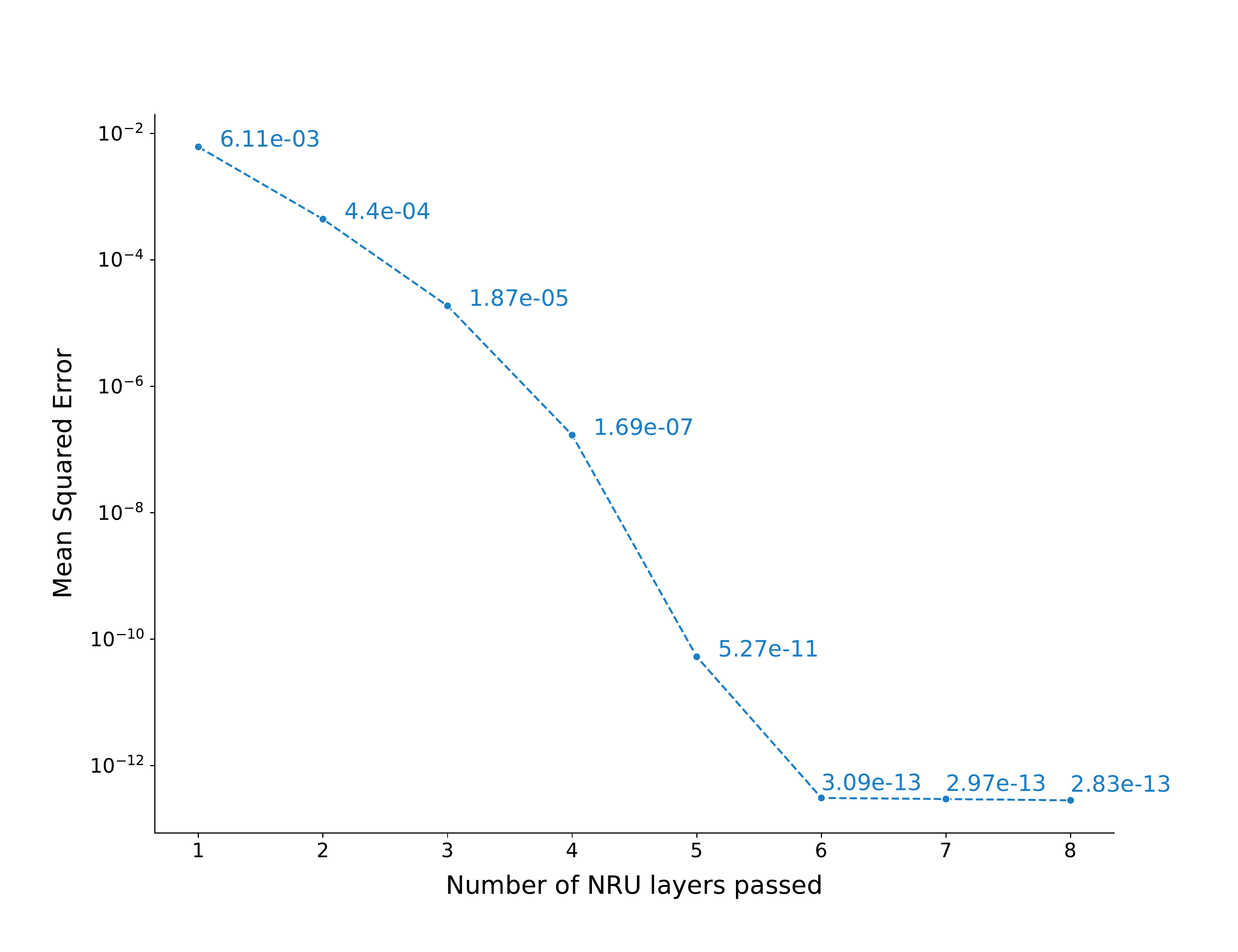}\caption{\label{fig:layer_number_effect}The degree of MSE change according to the number of NRU layers passed.}
	\end{figure}
	
	Figure~\ref{fig:layer_number_effect} depicts how the MSE changes each time it passes through the NRU layer of the NR emulation network. The MSE decreases by about $10^{-1}$ from the first to third NRU layers and by about $10^{-2}$ from the fourth to sixth NRU layers. In contrast, the seventh and eighth NRU layers reduce the MSE only slightly because the sixth layer has already virtually achieved the maximum possible accuracy of the single-precision floating number system. 
	
	\begin{table}[h]
		\begin{centering}
			\begin{tabular}{ccc}
				\hline 
				& \multicolumn{2}{c}{$\sigma=0.3$, $\tau=1$}\tabularnewline
				\cline{2-3} \cline{3-3} 
				\# of NRU layers passed & $k=1.5$ & $k=1.3$\tabularnewline
				\hline 
				\hline 
				0 & 0.90051656961441 & 0.72438144683838\tabularnewline
				1 & 0.37598699331284 & 0.32452529668808\tabularnewline
				2 & 0.30990260839462 & 0.30062055587769\tabularnewline
				3 & 0.30027109384537 & 0.30000048875809\tabularnewline
				4 & 0.30000036954880 & 0.30000001192093\tabularnewline
				5 & 0.30000007152557 & 0.30000001192093\tabularnewline
				6 & 0.30000016093254 & 0.30000001192093\tabularnewline
				7 & 0.30000001192093 & 0.30000001192093\tabularnewline
				8 & 0.30000001192093 & 0.30000001192093\tabularnewline
				\hline 
			\end{tabular}
			\par\end{centering}
		\centering{}\caption{\label{tab:two_sample}Change in the predicted value of the NR neural network according to the number of NRU layers passed.}
	\end{table}
	
	Last, we demonstrate how the inference value $\sigma_{pred}$ changes while passing through the NRU layers. Table~\ref{tab:two_sample} presents two specific cases: (1) $\sigma=0.3$, $\tau=1$, and $k=1.5$ and (2) $\sigma=0.3$, $\tau=1$, and $k=1.3$. The emulation network produces virtually the exact outputs at the seventh and fourth layers for $k=1.5$ and $k=1.3$, respectively. The outputs are indistinguishable from the exact implied volatility $\sigma_{impl}$ on the single-precision floating number system. These results confirm that the number of NRU layers required to achieve accurate implied volatility differs individually depending on the option.
	
	\section{Conclusion}
	Implied volatility is critical indicator that reflects expectations about future volatility and can be obtained by solving a nonlinear equation using the NR method. However, it is often necessary to repeatedly estimate numerous implied volatilities. The iterative method then fails because of a heavy computational burden. Therefore, the NR emulation network is proposed in this study to resolve the challenge. To develop the network, we implemented the NR method, like a PyTorch network, and optimized the network with TensorRT. As a result, the emulation network is up to 1,000 times faster than the NR method in SciPy. 
	
	Furthermore, the purpose of this work is achieved by emulating and optimizing the NR method without taking a complex mathematical approach. This result implies the possibility of solving other difficult issues of computational finance due to the recent progress in computing technology. Therefore, follow-up studies are required to address these problems using the neural emulation technique. 
	
	\section*{Acknowledgment}
	This research was supported by the BK21 Fostering Outstanding Universities for Research (No .5120200913674) funded by the Ministry of Education (Korea) and the National Research Foundation of Korea. Jeonggyu Huh received financial support from the National Research Foundation of Korea (Grant No. NRF-2022R1F1A1063371). This work was supported by the artificial intelligence industrial convergence cluster development project funded by the Ministry of Science and ICT (Korea) and Gwangju Metropolitan City.
	\bibliographystyle{elsarticle-harv}
	\bibliography{ref}
\end{document}